\documentstyle[twocolumn,psfig,fleqn,amstex,rotating]{mn}
\def\lsim{~\rlap{\raise 0.4ex\hbox{$<$}}{\lower 0.7ex\hbox{$\sim$}}~}
\def\gsim{~\rlap{\raise 0.4ex\hbox{$>$}}{\lower 0.7ex\hbox{$\sim$}}~}
\def\dd{{\rm d}}

\def\ts{T_{\rm s}}

\def\l0{L_\ast(0)}

\def\s0{S_\ast(0)}

\def\omg0{\Omega_0}
\def\dd{{\mathrm d}}

\def\a2{\alpha^{(2)}}

\def\ezero{\varepsilon_{_{0}}}
\def\eone{\varepsilon_{_{1}}}
\def\muzero{\mu_{_{0}}}
\def\muone{\mu_{_{1}}}
\def\pzero{p_{_{0}}}

\def\epsh{\eta_{_{\mathrm G}}}
\def\epsc{\eta_{_{\mathrm L}}}

\def\mpc3{\ {\rm Mpc^{-3}}}
\def\gpc3{\ {\rm Gpc^{-3}}}

\def\tkappa2{ {\tilde \kappa}^{2}}
\def\ts{{\tilde s}}
\def\tomega{{\tilde \omega}}
\def\cM{{\cal M}}
\def\cE{{\cal E}}

\begin{document}

\title[Disk instabilities with star formation]
{Instabilities of Galactic Disks  in the Presence of  Star Formation}
\author[Nusser]{Adi Nusser$^{1,2}$\\\\
$^{1}$Physics Department and the Asher Space Researrch Institute, Technion, Haifa 32000, Israel\\ $^2$Astrophysics, Oxford University, Keble Rd, Oxford OX1 3RK, UK\\}
\maketitle

\begin{abstract}
We discuss the stability of galactic disks in which 
the energy of interstellar clouds is gained  in encounters with expanding  
supernova remnants and lost in inelastic collisions.
Energy gain and loss processes introduce a phase difference between the 
pressure and density perturbations, making disks unstable on small scales for several recipes of star formation.  This is  in contrast to the standard stability analysis in which
small scale perturbations are stabilized by pressure. 
In the limit of small scales the  dispersion relation for the growth rate
 reduces to that 
of thermal instabilities in a fluid without gravity.
If instabilities lead to star formation, then 
our results imply a secondary  mode of star formation which operates on small
scales and  feeds on the existence of a primary 
mode on intermediate scales. This may be interpreted as a positive feedback. 
Further, the standard stability criterion on intermediate scales  is significantly  modified.
\end{abstract}



\section {Introduction}
\label{sec:introduction}

Instability and star formation in galactic disks are intimately linked. 
The observational support for that is the 
significant decline of star formation 
activity in regions with relatively low gas mass surface density,  as expected from various stability analyses
(e.g. Martin \& Kennicut  2001).
Theoretically,  
 the collapse of cold gas provides ripe conditions for the 
 appearance of molecular clouds and, subsequently,  star formation. 
(e.g. Spitzer 1968, Quirk 1972).

On large scales (of the order of the size of the disk) perturbations
are stabilized by rotation. Disks are also stable on small scales if 
the pressure, $P$, and density, $\rho$, are related by  $P\propto \rho^{\gamma}$, where $\gamma$ is a constant.
On  intermediate scales, 
neither 
rotation nor pressure can stop gravity from  amplifying the perturbations if the surface density is above a critical value. 
The first detailed stability calculations for  single fluid 
disks have been done in two classic papers  
by Toomre (1964) and Goldreich \& Lynden-Bell (1965).
Similar  type of analysis has been generalized to the more realistic 
case of 
disks containing stellar and gaseous components (Jog \& Solomon 1984, Rafikov 
2001, Griv, Gedalin \& Yuan 2002). Viscous disks have also been considered and found to be 
always unstable, with small scale perturbations growing  at a low rate that is set by the viscosity   (Gammie 1996).
These studies have assumed that the perturbation in the pressure is proportional
that in the density.

Here we consider the instability of the interstellar cloud component of the disk, when it is subject to local energy gain and loss processes.
Clouds gain energy in encounters with expanding supernova (SN) remnants, and
lose energy in inelastic collisions among each other (McKee \& Ostriker 1977, hereafter MO77). Both of these processes depend, in general, on the local 
surface density and velocity dispersion of the clouds.
The star formation rate is enhanced above the global average in a region with a positive density fluctuation, resulting in a local increase of the velocity 
dispersion. 
A perturbation in the velocity dispersion changes the energy loss rate and
may also affect the star formation rate. 
We perform detailed stability analysis under the assumption that the cloud component 
be treated as a hydrodynamical fluid over sufficiently large length and temporal scales.
The  treatment here leads to  
a significant modification of the standard stability criteria. 
 We get instabilities  on much smaller scales than what is inferred from the standard analysis. 
 In several important cases, instabilities extend down to the 
 scale  where the fluid treatment becomes invalid.

The  paper is organized as follows. 
The equations are derived and expanded in Laplace transforms in  \S\ref{sec:eqs}. Stability analysis is presented in \S\ref{sec:general}.
This section includes  a classification of stable and unstable modes by means of Nyquist diagrams, a discussion of a few limiting cases, and 
a treatment of instabilities with specific form for the energy gain and 
loss rates. 
The validity of our approach is assessed in 
 in \S\ref{sec:validity}. A summary and discussion of the results 
 are presented in \S\ref{sec:disc}. 

\section{The equations}
\label{sec:eqs}
We write the equations governing the evolution of perturbations in a thin
gaseous  disk. The gas represents 
interstellar clouds which, for simplicity, are assumed 
to have isotropic velocity dispersion. 
We assume that the gas has an ideal equation of state  which after integration over the height  of the disk yields
\begin{equation}
p=(\gamma-1)\mu\, \varepsilon 
\label{eq:eos} 
\end{equation}
to relate
the ``two-dimensional'' pressure, $p$, to the mass surface density, $\mu$,
and  the ``random'' kinetic  energy per unit mass, $\varepsilon=V_{\rm rms}^{2}/2$, where 
$V_{\rm rms}$ is the three dimensional velocity dispersion of the clouds, and $\gamma>1$. We will also refer to $\varepsilon$ as
internal energy. The adiabatic index $\gamma$ 
relates the projected quantities and, therefore, is in general different from 
the physical three dimensional index, $\Gamma$.  The relation between
$\gamma $ and $\Gamma$ depends on the structure of the disk. 
For a self gravitating disk $\gamma=3-2/\Gamma=9/5$ for $\Gamma=5/3$ (e.g. Gammie 2001). The difference between the two indices 
has little effect in our applications and so we present  numerical results only for $\gamma=5/3$. 
We write energy conservation in the following form,  
\begin{equation}
\frac{\dd \varepsilon}{\dd t}=\frac{p}{\mu^{2}}\frac{\dd \mu}{\dd t}+
G-L\; ,
\label{eq:econs}
\end{equation}
where $G(\mu,\varepsilon)$ and $L(\mu,\varepsilon)$ are, respectively, the energy gain and loss rates per unit mass, 
assumed to be explicit functions of $\mu$ and $\varepsilon$ alone. 
In the meantime we will study some general features of the  stability of the system without referring to specific forms for $G$ and $L$. 

Let the disk be in a steady state in which  
the disk is axisymmetric, gravity is balanced by rotation, and energy gain is balanced by 
energy loss, i.e.  $G=L$. This steady state is characterized by the 
rotational speed, $\Omega(r)$, the gas mass surface density, $\muzero(r)$,
 and the internal energy per unit mass, $\ezero(r)$, as functions of the distance from the center of the disk, $r$.
 We are interested in the linear response of the system to small 
perturbations from  this steady state. 
For simplicity we consider only radial perturbations as the complication
introduced by non-axisymmetric perturbations does not hide
any additional relevant physical effects. Let $u$, $v$, $\mu{^{(1)}}$, $\varepsilon^{(1)}$, $p^{(1)}$, and
$\phi^{(1)}$ be, respectively,  the perturbations in the radial velocity, the transverse velocity, the surface density, the internal energy,
the pressure, and the gravitational potential  field. 
The first order equations governing the evolution of these
perturbations are
\begin{eqnarray}
\label{eq:lin1}
 &p^{(1)}=(\gamma-1)\muzero \varepsilon^{(1)}+(\gamma-1)\ezero\mu^{(1)} \; ,&\\
\label{eq:lin2}
&\partial_{t}\varepsilon^{(1)}=\frac{p_{_{0}}}{\muzero^{2}}\partial_{t}\mu^{(1)}+C_{\mu}\mu^{(1)}+C_{\varepsilon}\varepsilon^{(1)}\; ,&\\
\label{eq:lin3} 
&\partial_{t}u-2\Omega(r)v +\partial_{r}p^{(1)}/\muzero(r)+\partial_{r}\phi^{(1)}\vert_{z=0}=0\; ,&\\
&\label{eq:lin4}\partial_{t} v -2B(r)u =0 \; ,&\\
&\label{eq:lin5}\partial_{t}\mu^{(1)}+\partial_{r}[r \mu_{_{0}}(r)u] =0\; ,&\\
&\label{eq:lin6}\partial_{r} (r\partial_{r}\phi^{(1)})/r
+\partial^{2}_{z}\phi^{(1)}=
4\pi G \mu_{_{1}}\delta^{\rm D}(z)\; ,&
\end{eqnarray}
where $p_{_{0}}=(\gamma-1)\ezero\muzero$, and
\begin{equation}
C_{\mu}=\frac{\partial(G-L)}{\partial \mu}\vert_{_{\muzero,\ezero}} \quad {\rm and} \quad
 C_{\varepsilon}=\frac{\partial(G-L)}{\partial \varepsilon}\vert_{_{\muzero,\ezero}}\; . 
 \label{eq:cdef}
\end{equation}
The first two of these equations are obtained by linearization 
of equations (\ref{eq:eos}) and(\ref{eq:econs}), respectively. 
The equations (\ref{eq:lin3}) and (\ref{eq:lin4}) are the radial and transverse   linear versions of 
the Euler equations. Mass conservation is represented by (\ref{eq:lin5}) and 
the Poisson equation by (\ref{eq:lin6}).
We  expand the perturbations in Fourier modes,
$\exp(i{\bf k \cdot r})$. Considering only modes satisfying 
$kr\gg 1$, the linear equations lead to, 
\begin{eqnarray}
&{\dot \varepsilon}_{_{1}}=\frac{\pzero}{\muzero^{2}}{\dot \mu}_{_{1}}+
C_{\mu}\muone+C_{\varepsilon}\eone &
\label{eq:mueps1} \\
&\nonumber \ddot{\mu}_{_{1}}+[\kappa^{2}-2\pi G\muzero k+(\gamma-1)\ezero k^{2 } ]      \muone&\\
&+(\gamma-1)\muzero k^{2}\eone=0\; ,&  
\label{eq:mueps2}
\end{eqnarray}
where $\muone$ is defined by $\mu^{(1)}(r,t)=\muone(t)\exp(i{\bf k \cdot r})$
and similarly for $\eone$, and
 $\kappa=2\Omega[1+(\dd \ln \Omega/\dd \ln r)/2]^{1/2}$ is the 
 epicyclic frequency.
In deriving these equations, the relation $\partial_{r}\phi^{(1)}|_{z=0}=i2\pi G \muone\exp(i{\bf k \cdot r})$ has been used (Toomre 1964).
 
 The differential equations (\ref{eq:mueps1}) and (\ref{eq:mueps2})  
are linear  with constant coefficients and they can be solved
by means of Laplace transformation (see Appendix A).
By taking the Laplace transform of  
(\ref{eq:mueps1})  and (\ref{eq:mueps2}) we get
\begin{eqnarray}
\nonumber&s\cE-\eone(0)=(\gamma-1)\frac{\ezero}{\muzero}(s\cM-\muone(0))&\\
&+C_{\mu}\cM+
C_{\varepsilon}\cE  \; ,&\\
&\nonumber s^{2}\cM-s\muone(0)-{\dot \mu}_{_{1}}(0)
+[\kappa^2-2\pi G\muzero k& \\&+(\gamma-1)\ezero k^2)]\cM+(\gamma-1)\muzero k^{2}\cE=0 \; ,&
\end{eqnarray}
where  $\cM(s)$ and $\cE(s)$ are the Laplace transforms of $\muone(t)$ and $\eone(t)$, respectively, and $\muone(0)$, ${\dot \mu}_{_{1}}(0)$, and $\eone(0)$ represent the 
initial conditions given at $t=0$. 
Solving  for $\cM $ yields, 
\begin{eqnarray}
B(s)\cM(s)&=&(s-C_{\varepsilon})[s\mu_{_{1}}(0)+{\dot \mu}_{_{1}}(0)]
\\
&+&(\gamma-1)\muzero k^{2}[\eone(0)-(\gamma-1)\frac{\ezero}{\muzero}\mu_{_{1}}(0)] \; , 
\end{eqnarray}
where
\begin{equation}
B(s)=s^{3}-s^{2}C_{\varepsilon}+s\omega_{0}^{2}-C_{\varepsilon}\omega_{1}^{2}\; ,
\label{eq:Bs}
\end{equation}
with
\begin{eqnarray}
\omega_{0}^{2}(k)&=&\kappa^{2}-2 \pi G\mu_{_{0}}k +\gamma(\gamma-1)
\ezero k^{2} \; , 
\label{eq:defom0} \\
\omega_{1}^{2}(k)&=&\omega_{0}^{2}-(\gamma-1)\ezero k^{2}\left(
\frac{\muzero C_{\mu}}{\ezero C_{\varepsilon}}+\gamma-1\right)\; .
\label{eq:defom1}
\end{eqnarray}
The third order polynomial $B(s)$ has at most three distinct roots, 
$s_{j}$ ($j=1,2,3$). Since $\cM(s)\propto 1/B(s)$ then 
according to the theory of Laplace transorms (see Appendix \ref{app:laplace})
$\muone(t)$ is a linear combination of $\exp(s_{j} t)$. 
Therefore, an unstable (growing) mode corresponds to 
a  root with a positive real part. 

In the limit of small scales, i.e. large $k$, 
the terms proportional to $k^{2}$ are dominant in (\ref{eq:defom0}) and
(\ref{eq:defom1}). This means that the effects of rotation and self-gravity
are negligible. In this limit, the characteristic equation $B(s)=0$, reduces 
to the dispersion relation derived by Field (1965) for the growth rate of 
thermal instabilities (see Eq. 15 in Field 1965, without the thermal conduction
term).

\section{Stability Analysis}
\label{sec:general}

Our task is to establish the relevant time scales for the evolution of 
the perturbations and classify the stable and unstable modes.
The three time scales corresponding to the roots 
$s_{j}$ of $B(s)$ must be reflected in $C_{\varepsilon}$, $\omega_{0}$ and $\omega_{1}$.
We will find that  $C_{\varepsilon}$ is negative for the specific forms 
of $G$ and $L$ we use below. 
The dependence on $C_{\varepsilon}$ can be scaled out by 
working with the variables 
 $\ts=s/(-C_{\varepsilon}) $, $\tomega_{0}=\omega_{0}/(-C_{\varepsilon})$
  and $\tomega_{1}=\omega_{1}/(-C_{\varepsilon})$. If ${\tilde s}_{j}$ ($j=1,2,3$)
 solves
 \begin{equation}
\ts^{3}+\ts^{2}+\ts\tomega_{0}^{2}+\tomega_{1}^{2}=0\; ,
\label{eq:tBs}
\end{equation}
then $s_{j}=-C_{\varepsilon}{\tilde s}_{j}$.
Therefore, the stability of a mode with a wave-number $k$, depends 
only on  $\omega_{0}^{2}(k)$ and $\omega_{1}^{2}(k)$. The time scales, however, 
may depend nontrivially on $C_{\varepsilon}$.

The roots of $B(s)$ can either be all real (with zero imaginary parts) 
or one real and two conjugate complex roots (i.e. having the same real part). 
   The sum of the roots is $s_{1}+s_{2}+s_{3}=C_{\varepsilon}<0$ and therefore 
   at least one of them must have a negative real part, meaning that there is
   always  a decaying mode. 
The question of the existence of unstable modes for 
any combination of $\omega_{0}^{2}$ and $\omega_{1}^{2}$ can be addressed by Nyquest diagrams
(Nyquist 1932, see also Appendix \ref{app:nyquist}).
These diagrams provide a simple way to determine the number of 
roots lying to right of the imaginary axis  (i.e. roots with  positive 
real parts) without having to solve the third order equation $B(s)=0$.
The outcome of this analysis is as follows (we refer only to roots lying to the 
right of the imaginary axis),
\begin{enumerate}
\item  $\omega_{0}^{2}>\omega_{1}^{2}>0\ $: no roots
\item  $\omega_{1}^{2}>\omega_{0}^{2}>0\ $:
2 complex roots
\item  $\omega_{0}^{2}>0$, $\omega_{1}^{2}<0$: 1 root 
\item $\omega_{0}^{2}<0$, $\omega_{1}^{2}<0$: 1 root
\item  $\omega_{0}^{2}<0$, $\omega_{1}^{2}>0$: 2 root (either 2 real or 2 complex conjugates) 
\end{enumerate}
Therefore, the disk 
is stable  only for $\omega_{0}^{2}>\omega_{1}^{2}>0$.  
\subsection{limiting cases}
Nyquest diagrams do not contain any information on the 
instability time scales. 
Those could be obtained by numerically finding  
the roots of $B(s)$.
Here we discuss  a few important limiting cases that can be studied analytically.  
 Let $\Delta \tomega^{2}=\tomega_{1}^{2}-\tomega_{0}^{2}$. 
First consider the limit, 
 $|\Delta \tomega^{2}|\ll |\tomega_{0}^{2}|$ in which the 
 roots differ by small amount from the values $\ts=-1$ and $\ts=\pm\sqrt{-\tomega^{2}}$ which solve (\ref{eq:tBs}) 
 with $\Delta \tomega^{2}=0$. In this case, 
\begin{eqnarray}
\nonumber s_{1}&=&-|C_{\varepsilon}|\frac{1+(\omega_1/C_\varepsilon)^2}{1+(\omega_0/C_\varepsilon)^{2}}\; ,\\
s_{2,3}&=&|C_{\varepsilon}|
\frac{\pm |C_{\varepsilon}|\sqrt{-\omega_{0}^{2}}+\omega_{0}^{2}}{C_{\varepsilon}^{2}+
\omega_{0}^{2}}\frac{\omega_{1}^{2}-\omega_{0}^{2}}{2\omega_{0}^{2}}\pm \sqrt{-\omega_{0}^{2}} \; . 
\end{eqnarray} 
Note that these expressions involve $C_{\varepsilon}$ in a complicated way. 
For $\omega_{1}^{2}>\omega_{0}^{2}>0$ the system is unstable, in accordance
with the Nyquist analysis, 
with a growth time scale of  $[{\rm Re}(s_{2})]^{-1}= 
[(1/2)C_{\varepsilon}(\omega_{1}^{2}-\omega_{0}^{2})/(C_{\varepsilon}^{2}+\omega_{0}^{2})]^{-1}$.
For $\omega_{0}^{2}>\omega_{1}^{2}>0$ there are no unstable modes.

The second and more important limit is when  $\tomega_{0}^{2}\gg 1$ and $\Delta \tomega^{2}={\cal O}(\omega_{0}^{2})$. 
This holds for 
small scale perturbations 
since for sufficiently large $k$ the leading  term in $\omega_{0}^{2}$ is 
the one involving $k^{2}$ (see eq.~\ref{eq:defom0}).
The roots of $B(s)$ are approximated by, 
\begin{equation}
 s_{1}=-|C_{\varepsilon}|\frac{\omega_{1}^{2}}{\omega_0^2} \; \quad {\rm and}\quad
 s_{2,3}=|C_{\varepsilon}|\frac{\omega_1^2-\omega_0^2}{2\omega_0^2}\pm i \omega_{0}\; . 
\label{eq:bigk}
\end{equation}
For $\omega_{1}^{2}>\omega_{0}^{2}$ the root $s_{1}$ represent 
a decaying mode, while the conjugate roots $s_{2}$ and $s_{3}$ 
correspond to  sound waves that grow at a rate proportional to  $|C_{\varepsilon}|$
and oscillate with frequency of $\omega_{0}/2\pi$.
The origin of the growth is  heating occuring during the  compression the waves (c.f. Field 1965).
In terms of $C_{\mu}$ and $C_{\varepsilon}$ the condition,    $\omega_{1}^{2}>\omega_{0}^{2}$, for the appearance  of growing sound waves reads, 
\begin{equation}
\nonumber \frac{\muzero C_{\mu}}{\ezero C_{\varepsilon}}+\gamma-1<0 \; ,  
\end{equation}
which is similar to the  condition in Eq.24 of Field 1965.  
For $\omega_{0}^{2}>\omega_{1}^{2}>0$ all modes are decaying with 
 $s_{2}$ and $s_{3}$ representing overstable (oscillating) modes.
For $\omega_{0}^{2}>0$ but $\omega_{1}^{2}<0$ 
the mode $s_{1}$ is growing without oscillations and $s_{2}$ and $s_{3}$ are overstable modes. The mode
described by $s_{1}$ corresponds to growth by condensation under
nearly constant pressure conditions as described by Field (1965).
It can result from an enhanced cooling efficiency 
 as the density is increased. 
The condition $\omega_{1}^{2}<0$ holds for
\begin{equation}
1-\frac{\muzero C_{\mu}}{\ezero C_{\varepsilon}}<0 \; .
\end{equation}
This is equivalent to the condition for a condensation mode 
as given by Eq.23 of Field 1965.

\subsection{Numerical solutions}
\begin{figure}
\hspace{-0.45cm}
\includegraphics[width=3.in]{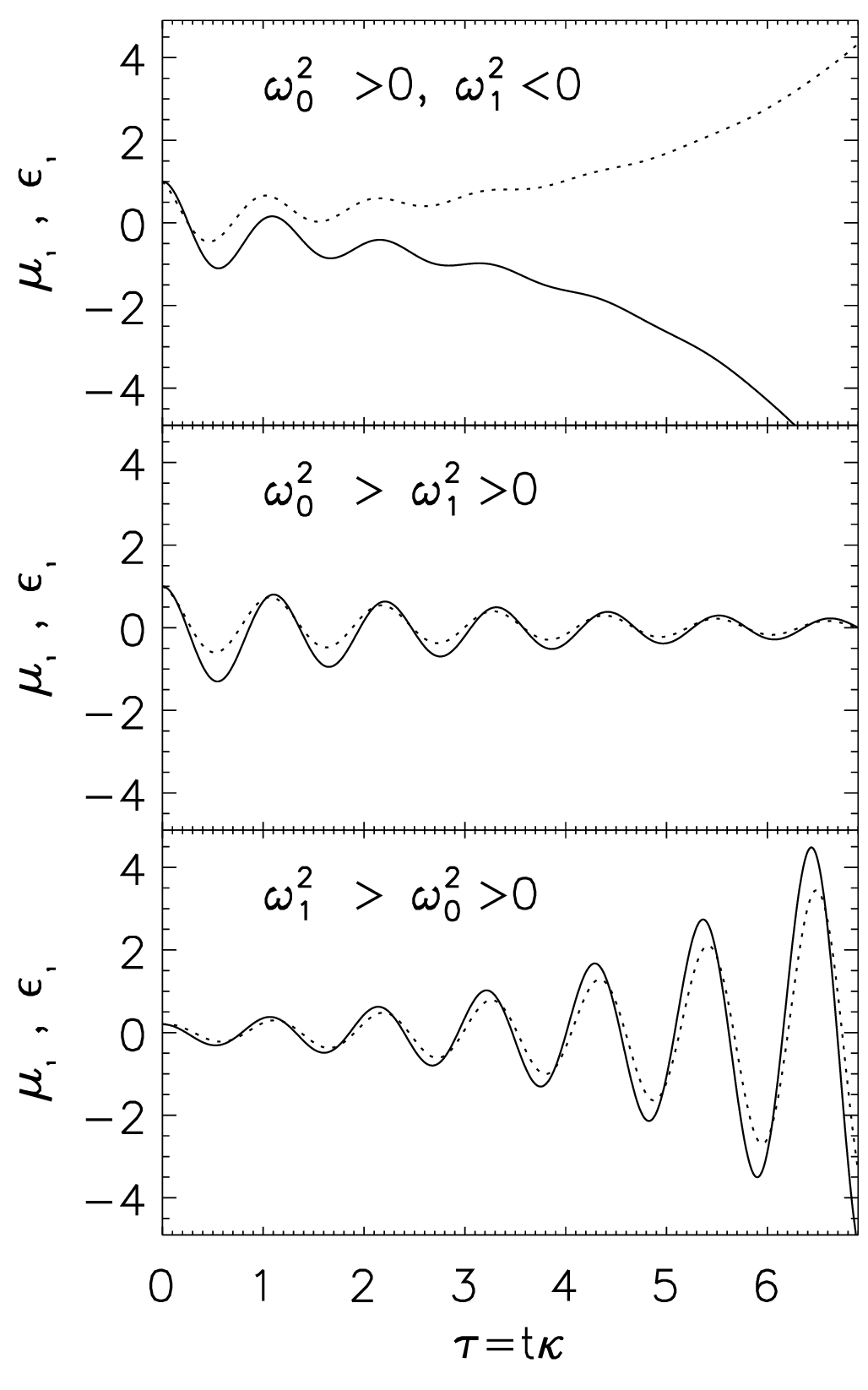}
\vspace{-0.1cm}
\caption{The time evolution of the perturbations in the surface density (solid line) and 
internal energy (dotted) in arbitrary units.
The curves have been obtained by direct numerical integration of the equations of motion. 
\label{fig:nsol}}
\end{figure}

To better visualize the time dependence  of perturbations we 
plot in figure (\ref{fig:nsol}) the functions $\muone(t)$ and 
$\eone(t)$  for $\omega_{1}^{2}<0$ (top panel), $\omega_{0}^{2}>\omega_{1}^{2}>0$ (middle) and $\omega_{1}^{2}>\omega_{0}^{2}$ (bottom). The figure shows only cases with
 $\omega_{0}^{2}>0$ which is always true for sufficiently large $k$.
 These curves 
have been obtained by numerical integration of the following dimensionless form of  
 (\ref{eq:mueps1}) and (\ref{eq:mueps2}),
\begin{eqnarray}
{{\tilde \mu}}_{_1}^{\prime\prime}+\left[1-\frac{2}{Q_0}\tilde  k 
+{\tilde k}^{2}\right]{\tilde \mu}_{_1}
+{\tilde k}^2\tilde {\varepsilon}_{_{1}}=0\; ,&&\\
{ {\tilde \varepsilon}}_{_{1}}^{\prime}=(\gamma-1)
{ {\tilde \mu}}_{_{1}}^{\prime}+{\tilde C}_{\mu} {\tilde \mu}_{_{1}}+{\tilde C}_{\varepsilon}{\tilde \varepsilon}_{_{1}}\; ,&&
\end{eqnarray}
 where the prime symbol denotes differential with respect to the variable $t\kappa$, $k_{0}=\pi G \muzero/((\gamma-1)\ezero Q_{0})$, $\tilde k=k/k_{0}$
 $Q_{0}=\kappa[(\gamma-1)\ezero]^{1/2}/(\pi G \muzero)$, 
 ${\tilde \varepsilon}_{_{1}}=\eone/\ezero$, ${\tilde \mu}_{_{1}}=\muone/\muzero$,  
 ${\tilde C}_{\mu}=\muzero C_{\mu}/(\ezero/\kappa)$, and  ${\tilde C}_{\varepsilon}=C_{\varepsilon} /\kappa$.  
 All solutions are for ${\tilde C}_{\varepsilon}=-1$, 
 ${\tilde k}=5$, $Q_{0}=1$, and $\gamma=5/3$. The solutions in the top, middle and bottom panels  correspond to ${\tilde C}_{\mu}=-1.3$,
 $0$, and $2$, respectively. The initial conditions of all
 solutions are ${{\tilde \mu}}_{_1}(0)={ {\tilde \varepsilon}}_{_{1}}(0)=1$ and $  {{\tilde \mu}}_{_1}^{\prime}=0$.
  The behavior of the solutions is in agreement with 
  Nyquest diagrams and the limiting cases. 
 For  $\omega_{1}^{2}>\omega_{0}^{2}>0$ (bottom panel), the solution is oscillatory with a growing 
 envelope. These growing sound waves
 arise because of the  slight heating during compression. The case $ \omega_{0}^{2}>\omega_{1}^{2}>0$ is overstable. 
 Note that the derivative of ${ {\tilde \varepsilon}}_{_{1}}$ at $\tau=0$  is positive although the dotted curve may appear as
 declining at $\tau=0$. The phase difference between the dotted and solid curves,  however, becomes more evident 
as $\tau$ increases.  

 For $\omega_{1}<0$ and $\omega_{0}^{2}>0$ the solution is unstable without oscillations. This mode is driven by thermal instability of a condensation
 mode (Field 1965).
 Note the phase difference between $\muone$ and $\eone$. This phase difference
 is the drive behind the instabilities shown in the top and bottom panels.

\section{Stability  with specific forms of $G$ and $L$}
\label{sec:specific}
A key parameter is 
the ratio $\muzero C_{\mu}/\ezero C_{\varepsilon}$ which 
determines  the difference between $\omega_{1}^{2}$ and 
$\omega_{0}^{2}$  (see Eq.\ref{eq:defom1}).
This ratio depends on the specific forms chosen to describe the 
energy gain and loss functions.   
 
 The function $L(\mu,\varepsilon)$ represents 
energy dissipation per units mass,  via inelastic cloud collisions. 
Hereafter we 
adopt the following form 
 (MO77, Efstathiou 2002),
\begin{equation}
L=\epsc \mu^{2}{\varepsilon}^{1/2}\; .
\label{eq:L}
\end{equation}

The energy gain function, $G$,  is directly proportional to the star formation rate
(MO77). 
We work with two distinct choices, the first is a Schmidt-Kennicutt (e.g. Kennicutt 1989) form  
and the second assumes that the star formation rate depends explicitly on the Toomre parameter
which governs instabilities on intermediate scales. 

\subsection{Stability with a Schmidt-Kennicutt  star formation law} 
Here we write the energy gain function as, 
\begin{equation}
G=\epsh \mu^{n} \; ,
\label{eq:kenn}
\end{equation}
where $\epsh $ is a numerical factor having the appropriate units.
Equating this $G$ with $L$ given by (\ref{eq:L}) gives,  
\begin{equation}
C_{\mu}= (n-2)\epsh\muzero^{n-1} \quad {\rm and}\quad
C_{\varepsilon}  = -\frac{1}{2}\frac{\epsc^{2}}{\epsh}\muzero^{4-n}\; .
\end{equation}
Therefore, using (\ref{eq:defom1}), 
\begin{equation}
\omega_{1}^{2}=\omega_{0}^{2}+(\gamma-1)(2n-3-\gamma)\ezero k^{2} \; .
\end{equation}


Of particular interest is the small scale behavior. 
For sufficiently large $k$ we write $\omega_{0}^{2}\approx \gamma (\gamma-1)\ezero k^{2}$. 
Therefore, for $n<3/2$, we  have $\omega_{1}^{2}<0$, while for $n>(3+\gamma)/2$, 
we have $\omega_{1}^{2}>\omega_{0}^{2}$. The disk is unstable in these two 
cases. But the nature of the instability is different:  the former is described by a single non-oscillatory mode,
while the latter has two growing oscillatory modes (see fig.~\ref{fig:nsol} and 
the discussion of Nyquist diagrams at the end  of \S\ref{sec:general}).

\begin{figure}
\hspace{-0.45cm}
\includegraphics[width=3.in]{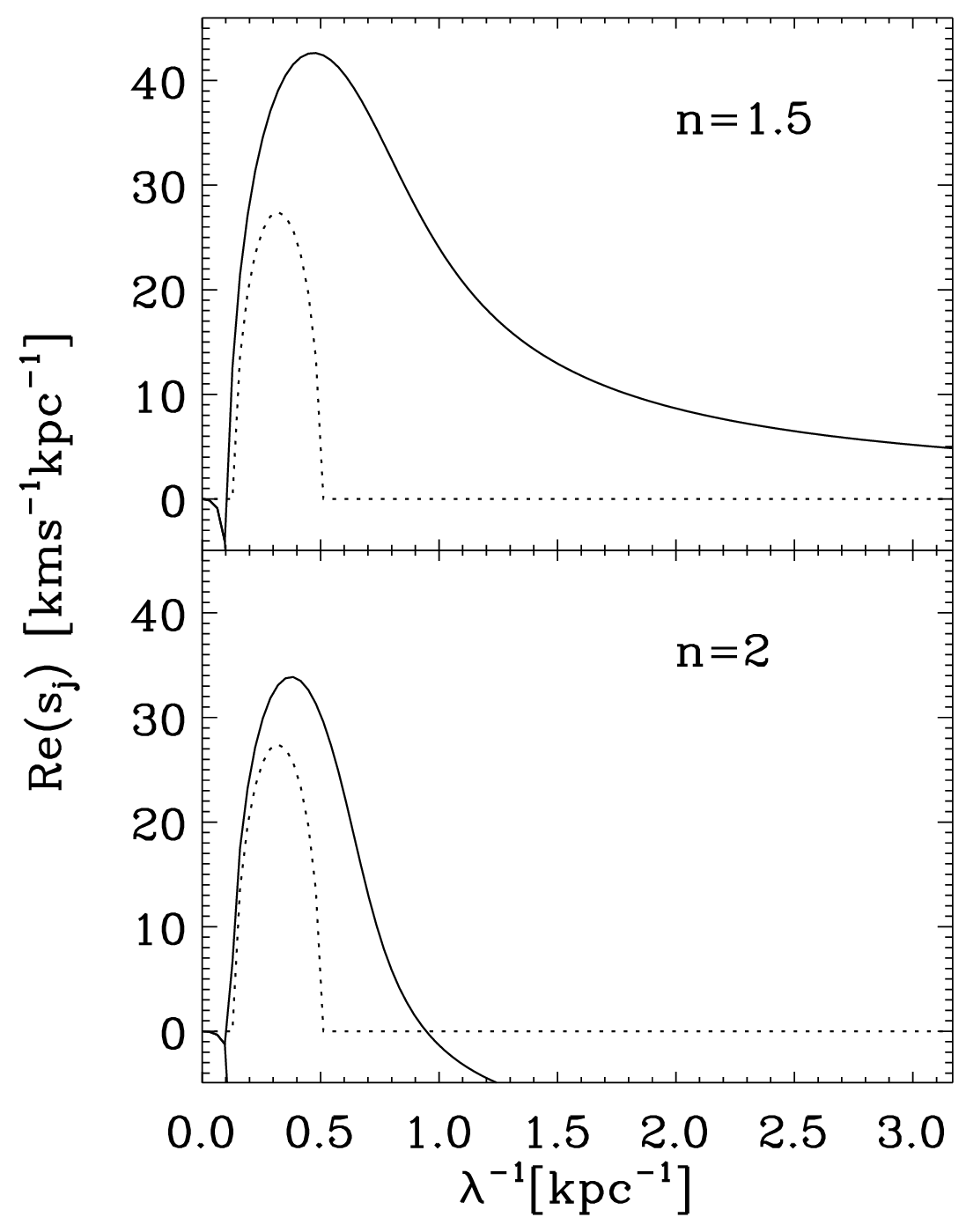}
\vspace{-0.1cm}
\caption{The growth rate of the unstable modes as  function of the scale $\lambda^{-1}=k/2\pi$. The top and bottom panels show results  for two values of the power index of the star formation rate,  $n$.
 The solid line is the real part of the root of $B(s)$. The dashed line is the growth rate estimated as $(-\omega_{0}^{2})^{1/2}$ for 
$\omega_{0}^{2}<0$ and zero otherwise. 
 The values of the various parameters were chosen such that the minimum 
 of $\omega_{0}^{2}$ matches the value given in Jog \& Solomon (1984). 
  \label{fig:roots}}
\end{figure}

We plot in figure (\ref{fig:roots})  the real parts of the roots as a function 
of the wave number $k$ for $n=1.5$ (top panel) and
$n=2$ (bottom). 
The dashed line is the instability growth rate
according to the  standard  analysis \`{a} la Toomre (1964). This  is given by  
$\sqrt{-\omega_{0}^{2}}$ for $\omega_{0}^{2}<0$, and zero 
otherwise.  
A value of $\gamma=5/3$ is used and $\kappa$, $\muzero$ and $\ezero$ have been  tuned so that the 
maximum of the dashed curve is consistent with the growth rate inferred from the lower curve of Fig.1a of Jog \& Solomon (1984). We use 
$|C_{\varepsilon}|=20 \; \rm km \; s^{-1}\;  kpc^{-1}$, corresponding to a time scale of 
$5\times 10^{7}\; \rm yr$ (MO77).
The solid line shows the positive real parts of the roots.
 According to the solid curve the system with $n=1.5$ remains unstable on small scales, albeit with
a longer time scale than the instabilities at intermediate scales. 
For $n=2$ the instability extends to smaller scales  than what would
be inferred from 
the standard analysis (dashed line).  

\subsection{$Q$ dependent star formation rate}
\begin{figure}
\hspace{-0.45cm}
\includegraphics[width=3.in]{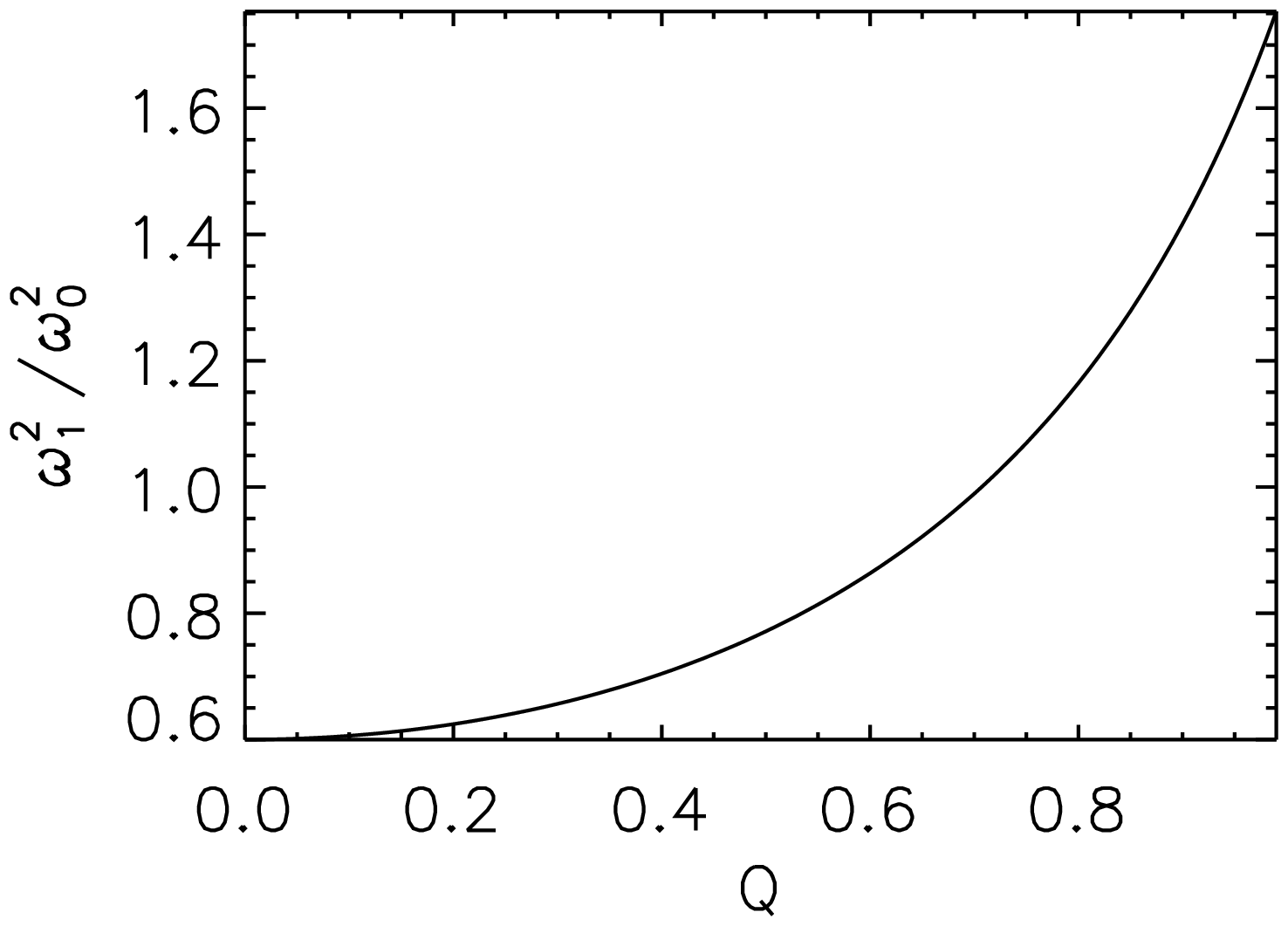}
\vspace{-0.1cm}
\caption{The ratio $\omega_{1}^{2}/\omega_{0}^{2}$ as a function
of the Toomre parameter for the Wang \& Silk star formation rate.
The ratio is computed in the limit $k\rightarrow \infty$. The system is unstable 
when the ratio exceeds unity.
\label{fig:omq}}
\end{figure}

Assume now that the star formation rate depends 
on the time scale defined by standard stability analysis (e.g. Toomre 1964).  
We work with a  gain function  of the form (e.g. Wang \& Silk 93, Elmegreen 1999), 
\begin{equation}
G=\epsh \kappa \mu F(Q)
\end{equation}
where  
\begin{equation}
Q=\frac{\kappa \varepsilon^{1/2}}{\pi G \mu} \quad
{\rm and} \quad F=\frac{(1-Q^{2})^{1/2}}{Q} \; .
\end{equation}
Note that the meaning of  $\epsh$ is different from (\ref{eq:kenn}); the same symbol  is used 
only for the sake of brevity.
Energy balance (G=L) at $\mu=\muzero$ and $\varepsilon=\ezero$ yields 
\begin{equation}
F(Q) =  Q/K \; ,
\end{equation}
where $1/K=(\frac{\epsc}{\epsh})(\frac{2\pi G} {\kappa^{2}})\muzero^{2}$. 
This relation gives,
\begin{equation}
Q^{2}=-\frac{K^{2}}{2}+\frac{K^{2}}{2}\left(1+\frac{4}{K^{2}}\right)^{1/2}
\; .
\end{equation}
 For  $K\ll 1$ and $ K\gg 1$ we have 
$Q=K^{1/2}$ and $Q=1$, respectively. 
The first order variation of $(G-L)$ 
is
\begin{eqnarray}
\delta G -\delta L=\epsh\kappa\left[F\muone+\mu\frac{\dd F}{\dd Q}\frac{\partial Q}{\partial \mu}\muone +\mu\frac{\dd F}{\dd Q}\frac{\partial Q}{\partial e}\eone\right]&&\\
-2\epsc\mu e^{1/2}\muone-\frac{1}{2}\epsc \mu^{2}e^{-1/2}\eone &&
\end{eqnarray}
Comparing this expression with the definitions of $C_{\mu}$ and $C_{\varepsilon}$ (see Eq.~\ref{eq:cdef}) and 
 using $\partial Q/\partial \mu=-Q/\mu$, 
$\partial Q/\partial e=Q/(2e)$ and  $\varepsilon^{1/2}=2\pi G\mu Q/\kappa $ we get, 
\begin{eqnarray}
C_{\mu}&=&-\epsh\kappa \left(1+\frac{\dd \ln F}{\dd \ln Q}\right) F\; , \\
C_{\varepsilon}&=&-\epsh\kappa\frac{\muzero}{2\ezero}
 \left(1-\frac{\dd \ln F}{\dd \ln Q}\right) F \; .
\end{eqnarray}
Since $\dd \ln F/\dd \ln Q<0$, the coefficient $C_{\varepsilon}$ is negative.
By evaluating the logarithmic derivative  in terms of $Q$ we find, $\mu C_{\mu}/\varepsilon C_{\varepsilon}=2Q^{2}/(Q^{2}-2)$, and 
\begin{equation}
\omega_{1}^{2}=\omega_{0}^{2}-(\gamma-1)\ezero k^{2}\left(\gamma-1+\frac{2Q^{2}}{Q^{2}-2}\right) \; . 
\end{equation}
For $Q^{2}>2(\gamma-1)/(\gamma+1)$, we have $\omega_{1}^{2}>\omega_{0}^{2}$, yielding 
unstable modes even on small scales.  For lower $q$, the system is stable on small scales but, as in the case with $n=2$ of the previous subsection,
unstable modes extend to scales smaller than in the standard analysis.   
The ratio $\omega_{1}^{2}/\omega_{0}^{2}$ as a function of $Q$ is plotted
in figure (\ref{fig:omq}), in the limit of very large $k$. A value of $\gamma=5/3$
is used in this figure.

\section{Validity of the approach}
\label{sec:validity}

The treatment of the clouds as a hydrodynamical fluid is valid  
 over scales larger than the mean free
path, $l_{\rm fp}$,  for cloud-cloud collisions. 
In the solar neighborhood $l_{\rm fp}\approx 100\; \rm pc$ for the cold neutral cores
and about 10 pc if the warm envelops are included (MO77). 
Although the warm envelops should contribute to the collisions, 
 we take  $l_{\rm fp}\approx 100\; \rm pc$ as an upper limit. 

Further, there are two time scale to be considered.
First, there is the mean time between collisions, $t_{\rm coll}=l_{\rm fp}/V_{\rm rms}\approx 3\times 10^{6}\; \rm yr$ for $V_{\rm rms}=10\; \rm km s^{-1}$.   
Second, there is  the mean time, $t_{\rm reheat}$, between two successive ``heating''
events of a cloud. This comes about because we assume that 
multiple encounters with  with expanding SN remnants eventually amount to an increase in random kinetic energy rather than bulk motions. 
According to Cox \& Smith (1974)  $t_{\rm reheat}\sim 10^{6}-10^{7}\; \rm yr$.

There is also the length   scale over which the 
velocity dispersion can equilibrate in a time $t={\rm max}(t_{\rm reheat},t_{\rm coll})$
This scale arises  
from the finite mean free path of the clouds (see also Gammie 1996) and is similar to the usual heat conduction (e.g. Zel'dovich \& Raizer 2002). 
We estimate this scale as $(t V_{\rm rms} l_{\rm fp})^{1/2}$.
Taking $t=t_{\rm coll}$
yields a scale of
$(t_{\rm coll} V_{\rm rms} l_{\rm fp})^{1/2}=l_{\rm fp}$.

Therefore, for an environment like the solar neighborhood, our approach 
is valid over scales larger than a hundred parsecs. 
The scale is sufficiently small that our treatment remains 
interesting. 
For example, this scale is smaller than the scale of  
 unstable modes found by means of standard stability analysis (e.g. Toomre 1964, Jog \& Solomon 1984). 

Finally, there is a continuous process of cloud destruction and production. We
assume that this process is rapid and always produces a fixed spectrum for 
the cloud size distribution.
 
\section{Summary and Discussion}
\label{sec:disc}

We have considered the instability of gaseous disks subject to 
local energy gain and loss processes in the presence of star formation. The gas represents interstellar clouds
and the energy is gained in repeated encounters of the clouds with 
expanding SN remnants and is lost in inelastic cloud-cloud collisions. 
These energy exchange processes introduce a phase difference between the
density and pressure perturbations. In several interesting situations 
this phase difference causes the pressure to amplify the density 
perturbation.
For a star formation rate proportional to $\mu^{n}$ 
the instabilities 
extend to much smaller scales than what is inferred from standard stability analyses 
\`{a} la Toomre (1964) and Goldreich \& Lynden-Bell (1965).

The small scale instabilities may be responsible for triggering further star formation.  Therefore, there may be two modes of star formation, a primary mode 
on intermediate scales
and a secondary mode operating on  small scales at a rate that is determined
by the coefficient $C_{\varepsilon}$ (see Eq.~\ref{eq:cdef}). This can be interpreted
as positive feedback in which  
the intermediate  scale mode is driving  star formation on smaller scales
through the development of instabilities. 
It is unclear what gain function one should 
use in (\ref{eq:econs}).   
A fully self-consistent stability analysis should incorporate   
the energy gain 
resulting from the  star formation mode that is induced by the instabilities.
This complicates   the problem substantially since 
the form of the gain function in this case must  be derived from the
stability analysis self-consistently. 
Nevertheless, we have found that  small scale instabilities 
develop at some level for most generic forms of the gain function.
We do not expect a full self-consistent treatment to change our conclusions
significantly.

We have not included the  coupling of gas to the stellar component.
This could easily be done, but does not affect the main conclusions 
of the present work. 
Including the stellar component 
would tend to destabilize small scale modes even further (Jog \& Solomon 1984).

\section*{Acknowledgment}
The author thanks Joe Silk for stimulating discussions, and an 
anonymous referee for pointing out an error in a previous version
of the paper and for several useful comments and clarifications.
 


\appendix
\section{A brief review of Laplace Transforms}
\label{app:laplace}

We briefly review
some of the relevant properties os Laplace Transforms  
The Laplace transform, $f(s)$, of a function $F(t)$, where $t\ge 0$,
is defined as 
\begin{equation}
f(s)\equiv {\cal L}\{F(t)\}=\int_0^\infty \exp\left(-st\right) F(t) \dd t .
\label{laplace:def}
\end{equation} 
We will need the Laplace transforms of first and second derivatives of a
function.  Using (\ref{laplace:def}) these transforms can be related to the
$f(s)$ by
\begin{eqnarray}
{\cal L} \{ F'(t)\} &=& s f(s)- F(0) \cr
{\cal L} \{ F''(t)\} &=& s^2 f(s)- s F(0) - F'(0) \; ,
\label{laplace:derv}
\end{eqnarray}
where the prime and double prime denote first and second order 
derivatives, respectively. 
The Bromwich integral expresses  $F(t)$ in terms of $f(s)$
as 
\begin{equation}
F(t)=\frac{1}{2\pi i}\int_{\gamma -i\infty}^{\gamma+i\infty}
\exp\left(s t\right) f(s)\dd s 
\label{bowich}
\end{equation}
where $i=\sqrt{-1}$ and $\gamma$ is a real number chosen so that
all  poles of $f(s)$ lie, in the complex plane, to the left of the 
vertical line defining the integration path. Therefore, by the residue
theorem we have
\begin{equation}
F(t)=\sum \left[
{\rm residues} \; {\rm of} \; \exp\left(s t\right)f(s)
\right]
\label{resid}
\end{equation}
As an example consider $f(s)=1/(s-s_1)(s-s_2)$ which has two simple poles at 
$s=s_1 $ and $s_2$. The residues of $\exp({s t}) f(s)$ at these poles are 
$\exp({s_1 t}) /(s_1-s_2)$ and $-\exp({s_2 t}) /(s_1-s_2)$ so that, by (\ref{resid}),
$F(t)=[\exp({s_1 t})- \exp({s_2 t})]/(s_1-s_2)$. If $s_1=s_2$,  the function 
has a pole of order two at $s_1$. The residue 
in this case is $\dd[(s-s_1)^2 \exp({st})f(s))]/\dd s$ evaluated at $s=s_1$. 
Therefore,  $F(t)=t \exp({s_1 t})$.

\section{Stability by means of Nyquist diagrams}
\label{app:nyquist}
Given a polynomial, $P_{N}$, of degree, $N$,  Nyquist diagrams provide an easy and elegant method for 
determining the number of its roots which lie to the right of the imaginary 
axis, i.e., roots with a positive real part. 
The method can be summarized as follows: 
$a)$ plot a contour of the 
value of the polynomial in the complex plane when its argument, $s$, 
is varied from 
$s=+i\infty$ to $s=-i\infty$ along the imaginary axis and  $b)$ use the contour diagram 
to compute the increase, $\Delta \psi$,  in the phase of 
$P_{N}(s)$ as $s$ moves along the imaginary axis. 
The number of roots with positive real parts is then $(\Delta \psi + N\pi)/(2\pi)$.
 
\begin{figure}
\hspace{-0.45cm}
\includegraphics[width=3.5in]{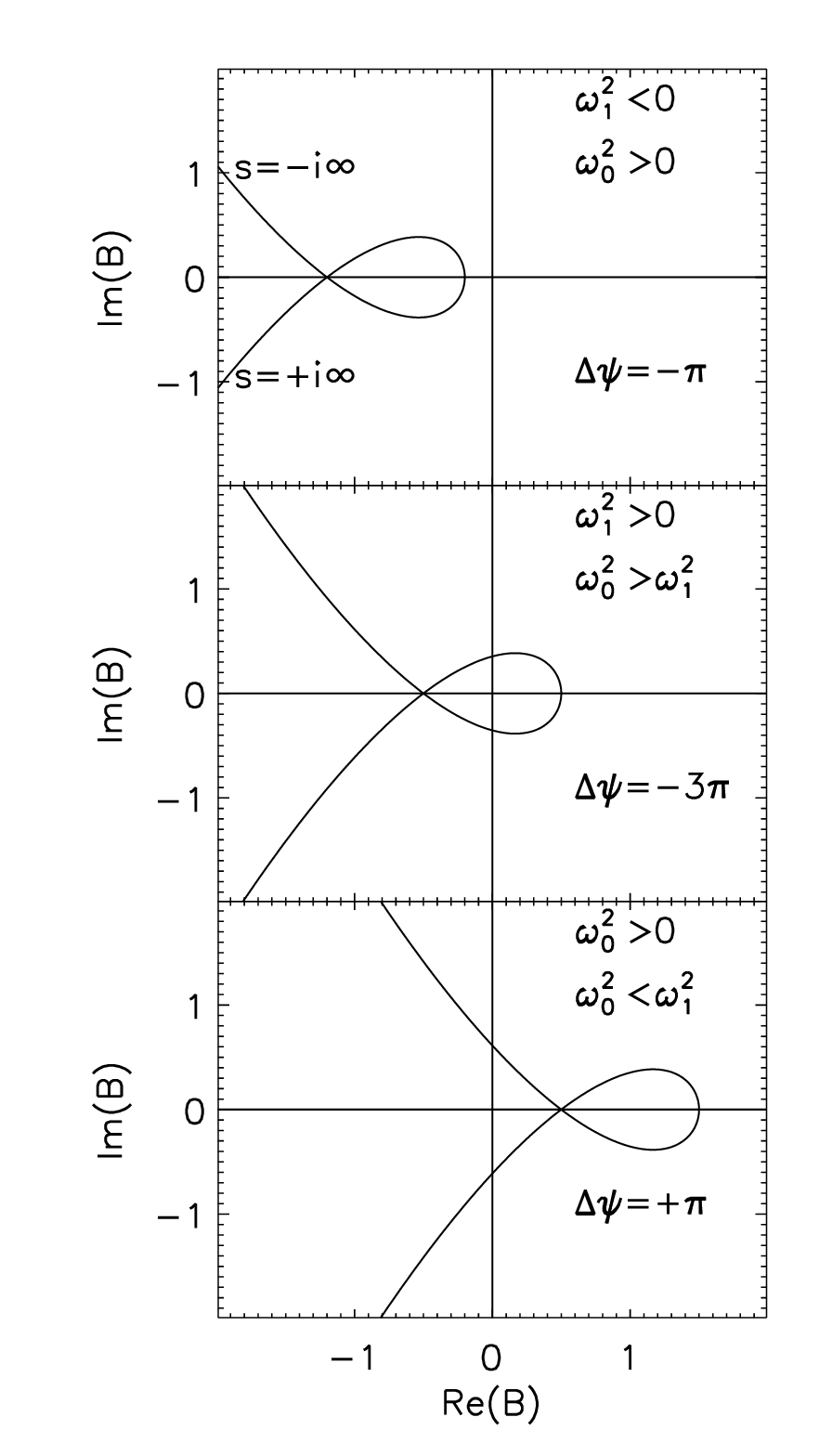}
\vspace{-0.1cm}
\caption{Nyquist diagrams of the polynomial $B(s)$. 
\label{fig:nyquist}}
\end{figure}

Figure (\ref{fig:nyquist}) shows Nyquist diagrams for the polynomial $B(s)$
given in (\ref{eq:Bs}) for three cases as indicated in the figure.
In the top panel the phase changes from $-\pi/2$ to $-3\pi/2$ and therefor
the number of roots with positive real parts is 1. Further, in this case this
root has zero imaginary part since otherwise its conjugate is also 
be a root and there would be two roots with positive (and equal) real parts. 
In the middle panel the change is $-3\pi$ and no roots lie to the 
right of the imaginary axis. 
In the bottom panel there are two roots with positive real parts.

\begin{thebibliography}{}

\bibitem{} Cox D.P., Smith B.W., 1974, ApJ, 189, L105
\bibitem{} Efstathiou G., 2000, MNRAS, 317, 697
\bibitem{} Elmegreen B.G., 1999, Proceedings of Star Formation 1999, Editor: Nakamoto T.
\bibitem{} Nyquist H., 1932, Bell. Syst. Technol. J., 11, 126, 139, 1217
\bibitem{} Field G., 1965, ApJ, 142, 531
\bibitem{} Gammie C.F., 1996, ApJ, 462, 725
\bibitem{} Gammie C.F., 2001, ApJ, 553, 174
\bibitem{} Goldreich P., Lynden-Bell D., 1965, MNRAS, 130, 97
\bibitem{} Griv E., Gedalin M., Yuan C., 2002, A\&A, 383, 338
\bibitem{} Hunter D.A., Elmegreen B.G., Baker A.L., 1998, ApJ, 493, 595
\bibitem{} Jog C.J.,   Solomon P.M., 1984, ApJ, 276, 114
\bibitem{} Kennicut R., 1989, ApJ, 344, 685
\bibitem{} Martin C.L., Kennicutt R.C., 2001, ApJ, 555, 301
\bibitem{} McKee C.F., Ostriker J.P., 1977, ApJ, 218, 148
\bibitem{} Quirk W.J., 1972, ApJ, 176, L9
\bibitem{} Rafikov R.R., 2001, MNRAS, 323, 445
\bibitem{} Spitzer L., 1968, Diffuse Matter in Space (New York: Wiley)
\bibitem{} Toomre A., 1964, ApJ, 139, 1217
\bibitem{} Wang B., Silk J., 1993, ApJ, 427, 759
\bibitem{} Zel'dovich Ya.B., Raizer Yu.P., 2002, Physics of Shock Waves and High-Temperature Hydrodynamic Phenomena (New York: Dover)

\end{thebibliography}
\end{document}